\documentclass[prl,twocolumn,nopacs,floatfix,amsmath,nofootinbib,amssymb,floatfix]{revtex4}
\usepackage{graphicx,color,dcolumn,booktabs,bm}
\usepackage{longtable,lscape}
\usepackage{txfonts}
\usepackage{overpic}
\usepackage{amssymb}
\usepackage{indentfirst}
\usepackage{feynmf}   
\usepackage{slashed}  
\usepackage{cases}
\usepackage{color}
\usepackage{multirow}
\usepackage{epstopdf}

\usepackage[colorlinks, citecolor=blue,anchorcolor=red,menucolor=red, linkcolor=red,filecolor=red,runcolor=red,urlcolor=blue,frenchlinks=red]{hyperref}

\graphicspath{{Figures/}} %

\begin{document}

\title{Predicting another doubly charmed molecular resonance  $T_{cc}^{\prime+}(3876)$}

\author{Rui Chen$^{1,2}$}\email{chenrui@hunnnu.edu.cn}
\author{Qi Huang$^{3,6}$}\email{huangqi@ucas.ac.cn}
\author{Xiang Liu$^{4,5,6}$}
\email{xiangliu@lzu.edu.cn}
\author{Shi-Lin Zhu$^{2}$}
\email{zhusl@pku.edu.cn}

\affiliation{ $^1$Key Laboratory of Low-Dimensional Quantum Structures and Quantum Control of Ministry of Education, Department of Physics and Synergetic Innovation Center for Quantum Effects and Applications, Hunan Normal University, Changsha 410081, China\\
$^2$Center of High Energy Physics, Peking University,
Beijing
100871, China\\
$^3$School of Physical Sciences, University of Chinese Academy of Sciences, Beijing 100049, China\\
$^4$School of Physical Science and Technology, Lanzhou University, Lanzhou 730000, China\\
$^5$Research Center for Hadron and CSR Physics, Lanzhou University and Institute of Modern Physics of CAS, Lanzhou 730000, China\\
$^6$Lanzhou Center for Theoretical Physics, Key Laboratory of
Theoretical Physics of Gansu Province, and Frontiers Science Center
for Rare Isotopes, Lanzhou University, Lanzhou 730000, China}

\begin{abstract}
The isospin breaking effect plays an essential role in generating
hadronic molecular states with a very tiny binding energy. Very
recently, the LHCb Collaboration observed a very narrow doubly
charmed tetraquark $T_{cc}^+$ in the $D^0D^0\pi$ mass spectrum,
which lies just below the $D^0D^{*+}$ threshold around 273 keV. In
this work, we study the $D^0D^{*+}/D^+D^{*0}$ interactions with the
one-boson-exchange effective potentials and consider the isospin
breaking effect carefully. We not only reproduce the mass of the
newly observed $T_{cc}^+$ very well in the doubly charmed molecular
tetraquark scenario, but also predict the other doubly charmed
partner resonance $T_{cc}^{\prime+}$ with $m=3876~\text{MeV}$, and
$\Gamma= 412~\text{keV}$. The prime decay modes of the
$T_{cc}^{\prime+}$ are $D^0D^+\gamma$ and $D^+D^0\pi^0$. The peculiar characteristic mass spectrum of the $D^0D^{*+}/D^+D^{*0}$ molecular systems  can be applied to identify the doubly charmed molecular states.

\end{abstract}

\maketitle

{\it Introduction---}As an important and effective approach to shed
light on the non-perturbative behavior of the quantum chromodynamics
(QCD), the study of the hadron spectroscopy has become an active
research field. Among abundant research issues around the hadron
spectroscopy, searching for the exotic hadronic matter is full of
challenges and opportunities at the birth of quark model
\cite{GellMann:1964nj,Zweig:1981pd,Zweig:1964jf}. There exist
different exotic hadronic matters like glueball, hybrid, multiquark
states.

Very recently, the LHCb Collaboration reported a very narrow state
$T_{cc}^+$ in the $D^0D^0\pi^+$ mass spectrum with the significance
over 10 $\sigma$~\cite{LHCb:2021vvq,LHCb:2021auc}. It is the firstly observed doubly
charmed tetraquark as its simplest valence quark component is
$cc\bar{u}\bar{d}$. Its spin-parity is assumed as $J^P=1^+$. Its
mass with respect to the $D^0D^{*+}$ threshold and width are
\begin{eqnarray}
\delta m &=& m_{T_{cc}^+}-(m_{D^0}+m_{D^{*+}})= -273\pm 61\pm 5^{+11}_{-14}~\text{keV}/c^2,\nonumber\\
\Gamma &=& 410\pm 165\pm 43^{+18}_{-38}~\text{keV},\nonumber
\end{eqnarray}
respectively. This important observation shall push the exploration
of exotic hadronic matter into a new era.

Since this doubly charmed tetraquark $T_{cc}^+$ locates just below
the $DD^*$ threshold, the doubly charmed meson-meson molecule is the
very attractive and probable explanation of the newly $T_{cc}^+$
state. The predicted mass of the $DD^*$ molecule with
$I(J^P)=0(1^+)$~\cite{Li:2012ss,Xu:2017tsr} is very consistent with
the mass of the newly $T_{cc}^+$. The binding energy is very small,
which is similar to the case of the $X(3872)$ as an isoscalar
$D\bar{D}^*$ molecule. The $T_{cc}^+$ and $X(3872)$ shares the same
one-pion-exchange force and nearly the same isospin breaking
pattern. The very tiny binding energy can amplify the isospin
breaking effect ~\cite{Li:2012cs}, which are also important for single heavy tetraquarks \cite{Gutsche:2016cml}. In fact, the interactions between
the $D^0D^{*+}$ and $D^+D^{*0}$ are almost the same. Therefore, if
the $T_{cc}^+$ is the $D^0D^{*+}$ molecule, there should exist the
other $D^+D^{*0}$ molecule just below the $D^+D^{*0}$ threshold.
When we recall the experimental information in the $D^0D^0\pi^+$
mass spectrum, one can find a minor structure existing between the
$D^0D^{*+}$ and $D^+D^{*0}$ threshold, which may correspond to the
$D^+D^{*0}$ bound state.

In this work, we perform a coupled channel analysis of the
$D^0D^{*+}/D^+D^{*0}$ interactions by adopting the OBE effective
potential and considering both the isospin breaking effect and $S-D$
wave mixing effect. We not only find that the newly $T_{cc}^+$ state
perfectly matches the very shallow bound $D^0D^{*+}/D^+D^{*0}$
molecular explanation with $J^P=1^+$, but also obtain a
coupled $T_{cc}^{\prime+}$ resonance with $m=3876~\text{MeV}$,
$\Gamma= 412~\text{keV}$. The $T_{cc}^+$ molecular state is mainly
composed of the $S-$wave $D^0D^{*+}$ component with over 70\%
probability. The dominant channel for the $T_{cc}^{\prime+}$
resonance is the $S-$wave $D^+D^{*0}$ component. The quantitative
analysis on the strong decay behavior for the $T_{cc}^{\prime+}$
resonance indicates the $D^+D^0\pi^0$ channel is more important than
the $D^0D^0\pi^+$, which explains why the experimental events for
the $T_{cc}^{\prime+}$ in the $D^0D^0\pi^+$ mass spectrum are not so
significant. Nevertheless, it is likely to search for the
$T_{cc}^{\prime+}$ state in the $D^0D^+\gamma$ and $D^+D^0\pi^0$
final states.

{\it $D^0D^{*+}/D^+D^{*0}$ interactions and the isospin breaking
effect---} In this work, a crucial input is the interaction between
the charmed mesons. So far, the understanding of the realistic
interactions is not enough due to the poor experimental data.
Several phenomenological models have been put forward, such as the
multiquark model, the QCD sum rule, the OBE model, the effective
field theory (see reviews
\cite{Chen:2016qju,Liu:2019zoy,Guo:2017jvc} for details. Especially
the $T_{cc}$ has been reviewed extensively in Ref.
\cite{Liu:2019zoy}). We adopt the OBE model to study the mass
spectrum of the $D^0D^{*+}/D^+D^{*0}$ system with $J^P=1^+$
and consider the $\pi/\sigma/\eta/\rho/\omega$ meson exchange
interactions.

The general procedure for deducing the OBE effective potentials is
organized as follows. After constructing the effective Lagrangians,
one can easily write down the OBE scattering amplitude
$\mathcal{M}[M_1{M}_2\to M_3{M}_4]$ for the $M_1{M}_2\to M_3{M}_4$
process. The OBE effective potentials in the momentum space can be
related to the corresponding scattering amplitudes by a Breit
approximation, $\mathcal{V}(\bm{q})=-\mathcal{M}[M_1{M}_2\to
M_3{M}_4]/\sqrt{16m_{M_1}m_{{M}_2}m_{M_3}m_{{M}_4}}$. In order to
obtain the OBE effective potentials in the coordinate space, we
further perform a Fourier transformation, $\mathcal{V}(\bm{r})
=\int\frac{d^3\bm{q}
e^{i\bm{q}\cdot\bm{r}}}{(2\pi)^3}\mathcal{V}(\bm{q})\mathcal{F}^2(q^2,m_E^2)$.
Here, we introduce a monopole form factor at every interaction
verteces to compensate the off-shell effect of the exchanged meson,
which has the form of
$\mathcal{F}^2(q^2,m_E^2)=(\Lambda^2-m_E^2)/(q^2-m_E^2)$. $\Lambda$,
$m_E$, and $q$ stand for the cutoff, mass and four-momentum of the
exchanged particle, respectively.

The relevant effective Lagrangians to describe the interactions
between the $S-$wave charmed mesons and the light mesons are
constructed in terms of heavy quark symmetry and chiral symmetry
\cite{Yan:1992gz,Wise:1992hn,Burdman:1992gh,Casalbuoni:1996pg,Falk:1992cx,Ding:2008gr},
i.e.,
\begin{eqnarray}
\mathcal{L}_{{\mathcal{P}}^{(*)}{\mathcal{P}}^{(*)}\sigma} &=& -2g_s{\mathcal{P}}_b^{\dag}{\mathcal{P}}_b\sigma-2g_s{\mathcal{P}}_b^{*}\cdot{\mathcal{P}}_b^{*\dag}\sigma,\\
\mathcal{L}_{{\mathcal{P}}^{(*)}{\mathcal{P}}^{(*)}\mathbb{P}} &=&
-\frac{2g}{f_{\pi}}\left({\mathcal{P}_b}{\mathcal{P}}_{a\lambda}^{*\dag}
+{\mathcal{P}}^*_{b\lambda}{\mathcal{P}}_{a}^{\dag}\right)\partial^{\lambda}\mathbb{P}_{ba}\nonumber\\
  &&-i\frac{2g}{f_{\pi}}v^{\alpha}\varepsilon_{\alpha\mu\nu\lambda}{\mathcal{P}}_b^{*\mu}{\mathcal{P}}_{a}^{*\lambda\dag}
\partial^{\nu}\mathbb{P}_{ba},\\
\mathcal{L}_{{\mathcal{P}}^{(*)}{\mathcal{P}}^{(*)}\mathbb{V}} &=& -\sqrt{2}\beta g_V{\mathcal{P}}_b\cdot{\mathcal{P}}_a^{\dag} v\cdot\mathbb{V}_{ba}\nonumber\\
     &&-2\sqrt{2}\lambda g_Vv^{\lambda}\varepsilon_{\lambda\mu\alpha\beta}\left({\mathcal{P}}_b{\mathcal{P}}^{*\mu\dag}_a
     +{\mathcal{P}}_b^{*\mu}{\mathcal{P}}_a^{\dag}\right)\partial^{\alpha}\mathbb{V}^{\beta}_{ba}\nonumber\\
    &&-\sqrt{2}\beta g_V{\mathcal{P}}_b^*\cdot{\mathcal{P}}_a^{*\dag}v\cdot\mathbb{V}_{ba}\nonumber\\
   &&-i2\sqrt{2}\lambda g_V{\mathcal{P}}_b^{*\mu}{\mathcal{P}}_a^{*\nu\dag}
   \left(\partial_{\mu}\mathbb{V}_{\nu}-\partial_{\nu}\mathbb{V}_{\mu}\right)_{ba}
\end{eqnarray}
with the pseudoscalar mesons fields $\mathcal{P}^T=(D^+,~D^0)$ and
vector mesons fields $\mathcal{P}^{*T}=(D^{*+},~D^{*0})$.
$v=(1,\textbf{0})$. The light pseudoscalar meson matrix $\mathbb{P}$
and the light vector meson matrix $\mathbb{V}_{\mu}$ have the
conventional forms of
\begin{eqnarray}
\left.\begin{array}{c} {\mathbb{P}} = {\left(\begin{array}{ccc}
       \frac{\pi^0}{\sqrt{2}}+\frac{\eta}{\sqrt{6}} &\pi^+ &K^+\\
       \pi^-       &-\frac{\pi^0}{\sqrt{2}}+\frac{\eta}{\sqrt{6}} &K^0\\
       K^-         &\bar K^0   &-\sqrt{\frac{2}{3}} \eta     \end{array}\right)},\\
{\mathbb{V}}_{\mu} = {\left(\begin{array}{ccc}
       \frac{\rho^0}{\sqrt{2}}+\frac{\omega}{\sqrt{2}} &\rho^+ &K^{*+}\\
       \rho^-       &-\frac{\rho^0}{\sqrt{2}}+\frac{\omega}{\sqrt{2}} &K^{*0}\\
       K^{*-}         &\bar K^{*0}   & \phi     \end{array}\right)}_{\mu},
\end{array}\right.
\end{eqnarray}
respectively. We employ the coupling constants $g_s=0.76$
\cite{Machleidt:1987hj,Wang:2019nwt}, $g=0.59$
\cite{Isola:2003fh,Zyla:2020zbs}, $f_{\pi}=0.132$ GeV, $\beta=0.9$
\cite{Isola:2003fh}, $\lambda=0.56~\text{GeV}^{-1}$
\cite{Isola:2003fh}, and $g_V=5.8$.

When we consider the isospin breaking effect and the $S-D$ wave
mixing effect, the discussed channels for the coupled
$D^0D^{*+}/D^+D^{*0}$ systems include $D^0D^{*+}({}^3S_1)$,
$D^0D^{*+}({}^3D_1)$, $D^+D^{*0}({}^3S_1)$, and
$D^+D^{*0}({}^3D_1)$. The OBE effective
potentials for the coupled $D^0{D}^{*+}/D^+D^{*0}$ system
with $J^P=1^+$ read as
\begin{eqnarray}
V=\left(\begin{array}{cc}\mathcal{V}^{D^0{D}^{*+}\to D^0D^{*+}}      &\mathcal{V}^{D^+{D}^{*0}\to D^0D^{*+}}\\
        \mathcal{V}^{D^0{D}^{*+}\to D^+D^{*0}}       &\mathcal{V}^{D^+{D}^{*0}\to D^+D^{*0}}\end{array}\right),
\end{eqnarray}
with
\begin{eqnarray}
\mathcal{V}^{D^0{D}^{*+}\to D^0D^{*+}} &=&
-g^2_{s}\mathcal{Y}_{\sigma}
    +\frac{g^2}{3f^2_{\pi}}\mathcal{Z}_{\pi0}^{\prime}
    -\frac{1}{4}\beta^2g_V^2\left(\mathcal{Y}_{\rho}-\mathcal{Y}_{\omega}\right)\nonumber\\
    &&+\frac{2}{3}\lambda^2g^2_V\mathcal{X}_{\rho0},\label{eff1}\\
\mathcal{V}^{D^+{D}^{*0}\to D^0D^{*+}} &=&
-\frac{g^2}{6f^2_{\pi}}\left(\mathcal{Z}_{\pi2}-\frac{1}{3}\mathcal{Z}_{\eta2}\right)
    +\frac{1}{2}\beta^2g_V^2\mathcal{Y}_{\rho}\nonumber\\
    &&-\frac{1}{3}\lambda^2g^2_V\left(\mathcal{X}_{\rho2}-\mathcal{X}_{\omega2}\right),\\
\mathcal{V}^{D^+{D}^{*0}\to D^+D^{*0}} &=&
-g^2_{s}\mathcal{Y}_{\sigma}
    +\frac{g^2}{3f^2_{\pi}}\mathcal{Z}_{\pi1}
    -\frac{1}{4}\beta^2g_V^2\left(\mathcal{Y}_{\rho}-\mathcal{Y}_{\omega}\right)\nonumber\\
    &&+\frac{2}{3}\lambda^2g^2_V\mathcal{X}_{\rho1}.\label{eff2}
\end{eqnarray}
The variables in the above equations are
\begin{eqnarray*}
\left.\begin{array}{ll}  q_0=m_{D^{*+}}-m_{D^0},    &q_1=m_{D^{*0}}-m_{D^+}, \\   q_2=\frac{m_{D^{*+}}^2+m_{D^+}^2-m_{D^0}^2-m_{D^{*0}}^2}{2(m_{D^0}+m_{D^{*+}})},\quad\quad\quad\\
\Lambda_i^2=\Lambda^2-q_i^2,   &m_{\rho/\omega i}^2=m_{\rho/\omega}^2-q_i^2,\\
m_{\pi0}^2=q_0^2-m_{\pi^+}^2,    &m_{\pi1}=m_{\pi^+}^2-q_1^2,\\
m_{\pi2}=m_{\pi^0}^2-q_2^2,      &m_{\eta2}=m_{\eta}^2-q_2^2.
\end{array}\right.
\end{eqnarray*}

In the above expressions, we have defined several useful functions,
\begin{eqnarray*}
\mathcal{Y}_E &=& \mathcal{O}_1[J^P]Y(\Lambda,m_E,r)\nonumber\\
     &=& \mathcal{O}_1[J^P]\left(\frac{e^{-m_E r}-e^{-\Lambda^2 r}}{4\pi r}-\frac{\Lambda^2-m^2_E}{8\pi\Lambda}e^{-\Lambda r}\right),\\
\mathcal{Z}_{Ei} &=& \left(\mathcal{O}_1[J^P]\nabla^2
     +\mathcal{O}_2[J^P]r\frac{\partial}{\partial r}\frac{1}{r}\frac{\partial}{\partial r}\right)Y(\Lambda_i,m_{Ei},r),\\
\mathcal{Z}_{Ei}^{\prime} &=& \left(\mathcal{O}_1[J^P]\nabla^2
     +\mathcal{O}_2[J^P]r\frac{\partial}{\partial r}\frac{1}{r}\frac{\partial}{\partial r}\right)\nonumber\\
    &&\times\left(\frac{\cos (m_{Ei} r)-e^{-\Lambda_i  r}}{4 \pi  r}-\frac{\left(\Lambda_i^2+m_{Ei}^2\right) \exp (-\Lambda_i  r)}{8 \pi \Lambda_i }\right),\\
\mathcal{X}_{Ei} &=& \left(2\mathcal{O}_1[J^P]\nabla^2
     -\mathcal{O}_2[J^P]r\frac{\partial}{\partial r}\frac{1}{r}\frac{\partial}{\partial r}\right)Y(\Lambda_i,m_{Ei},r).
\end{eqnarray*}
Here, $\mathcal{O}_1[J^P]$ and $\mathcal{O}_2[J^P]$ are the
spin-spin interactions $\bm{\epsilon}_1\cdot\bm{\epsilon}_3^{\dag}$,
$\bm{\epsilon}_1\cdot\bm{\epsilon}_4^{\dag}$ and the tensor force
operators
$S\left(\hat{r},\bm{\epsilon}_1,\bm{\epsilon}_3^{\dag}\right)$,
$S\left(\hat{r},\bm{\epsilon}_1,\bm{\epsilon}_4^{\dag}\right)$ with
$S(\hat{r}, \bm{x},
\bm{y})=3(\hat{r}\cdot\bm{x})(\hat{r}\cdot\bm{y})-(\bm{x}\cdot\bm{y})$,
respectively. In our calculations, these operators are replaced by
numerical matrices $\langle f|\mathcal{O}|i\rangle$, $|i\rangle$ and
$\langle f|$ stand for the spin-orbit wave functions for the initial
and final states, respectively, i.e.,
\begin{eqnarray}
\left.\begin{array}{c}
\bm{\epsilon}_1\cdot\bm{\epsilon}_3^{\dag}\\\bm{\epsilon}_1\cdot\bm{\epsilon}_4^{\dag}\end{array}\right.\mapsto
\left(\begin{array}{cc}1  &0\\    0   &1\end{array}\right),\quad
\left.\begin{array}{c}
S\left(\hat{r},\bm{\epsilon}_1,\bm{\epsilon}_3^{\dag}\right)\\S\left(\hat{r},\bm{\epsilon}_1,\bm{\epsilon}_4^{\dag}\right)\end{array}\right.\mapsto
\left(\begin{array}{cc}0  &-\sqrt{2}\\    -\sqrt{2}
&1\end{array}\right).
\end{eqnarray}
Here, we can also find an
approximate relation between the OPE effective potentials, like
$V_{\pi}^{D^+D^{*0}\to D^0D^{*+}}\sim -2V_{\pi}^{D^0D^{*+}\to
D^0D^{*+}}$, where the slight deviation from the exact relation is
due to the different exchanged momentum of the exchanged mesons.

\begin{figure}[!htbp]
\centering
\includegraphics[width=3.3in]{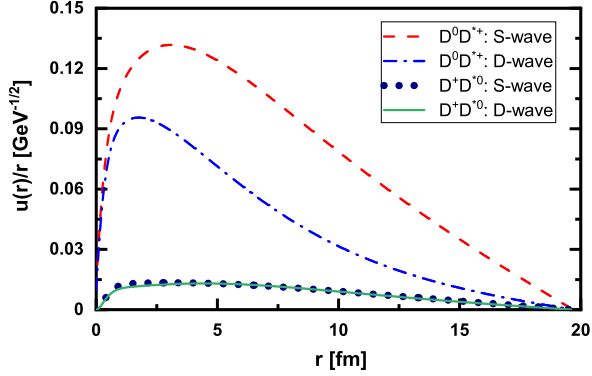}
\caption{The radial wave functions for the doubly charmed
$D^0D^{*+}/D^+D^{*0}$ molecular state with $J^P=1^+$.}
\label{wave}
\end{figure}

We first solve the coupled channel Shr$\ddot{\text{o}}$dinger
equation and obtain the bound state properties for the coupled
$D^0D^{*+}/D^+D^{*0}$ system with $J^P=1^+$. When the cutoff
takes a reasonable value $\Lambda=1.16$ GeV, we find a very shallow
bound state with a binding energy $E=-259.90$ keV. Thus, its mass
$M=m_{D^0}+m_{D^{*+}}+E$ overlaps with the mass of the $T_{cc}^+$.
In Figure \ref{wave}, we present the radial wave functions for the
doubly charmed $D^0D^{*+}/D^+D^{*0}$ molecular state with
$J^P=1^+$. It is a typical loosely bound molecular state. Its
root-mean-square radius is 6.28 fm, which is much larger than the
size of its components. The $S-$wave $D^0D^{*+}$ and $D^+D^{*0}$
components are dominant. Their probabilities $\int
d^3r|\psi_i(r)|^2/\sum_i\int d^3r|\psi_i(r)|^2$ are 72.52\% and
25.85\%, respectively.

\begin{figure}[!htbp]
\centering
\includegraphics[width=3.1in]{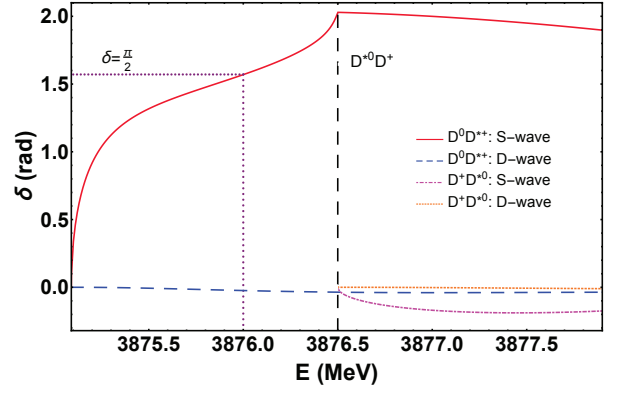}
\caption{Phase shifts for the coupled $D^0D^{*+}/D^+D^{*0}$ system
with $J^P=1^+$. Here, we adopt the same cutoff value
$\Lambda=1.16$ GeV, the dotted line shows the mass position of the
obtained doubly charmed resonance.} \label{phase}
\end{figure}

We further study the phase shifts for the coupled
$D^0D^{*+}/D^+D^{*0}$ systems with $J^P=1^+$ to search for the
doubly charmed resonant tetraquark state. Here, we adopt the same
OBE effective potentials for the coupled $D^0D^{*+}/D^+D^{*0}$
systems with $J^P=1^+$ and the same cutoff value
$\Lambda=1.16$ GeV. In general, a typical Breit-Wigner resonance
appears in the position with $\delta=(n+1/2)\pi$, where the cross
section $\sigma(E)$ reaches the maximum $\sigma_{\text{Max}}(E_0)$,
and $E_0$ corresponds to the mass of the resonance. The width of the
resonance reads as
$\Gamma=2/\left(\frac{d\delta(E)}{{d}E}\right)_{E_0}$. As shown in
the Figure \ref{phase}, there exists a doubly charmed resonance
$T_{cc}^{\prime+}$ in the phase shift for the $S-$ wave $D^0D^{*+}$
channel, its mass and width are
\begin{eqnarray}
m=3876~\text{MeV},\quad  \Gamma= 412~\text{keV},
\end{eqnarray}
respectively. In fact, the $T_{cc}^{\prime+}$ state is not a
shape-type resonance but a Feshbach-type resonance. Once we turn off
the contribution from the $D^+D^{*0}$ channel, it disappears. Thus,
the isospin breaking effect plays a very important role in forming
the $T_{cc}^{\prime+}$ state.

\begin{figure}[!htbp]
\includegraphics[width=3.3in]{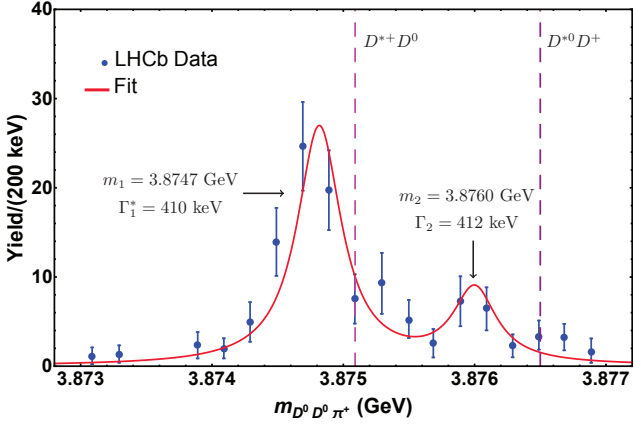}
\caption{The fit to the experimental data of the $D^0D^0\pi^+$ mass
spectrum through the obtained masses and width for the doubly
molecule and the doubly resonance. Here, $\Gamma_1^*=410$ keV is the
experimental width of the newly $T_{cc}^+$~\cite{LHCb:2021vvq}. The dash
lines label the $D^0D^{*+}$ and $D^+D^{*0}$ mass thresholds.}
\label{mass}
\end{figure}

To summarize, we have obtained a loosely bound doubly charmed
molecular tetraquark and a doubly charmed resonance from the
$D^0D^{*+}/D^+D^{*0}$ interactions with $J^P=1^+$. In Figure
\ref{mass}, we fit the $D^0D^0\pi^+$ mass spectrum with our obtained
mass for the $T_{cc}^{+}$ molecule state and our obtained mass
and width for the $T_{cc}^{\prime+}$ resonance. We use the
experimental value $\Gamma=410$ keV for the decay width of the
$T_{cc}^+$ in the fitting~\cite{LHCb:2021vvq}. Our fit is consistent
with the experimental data. Therefore, after considering the isospin
breaking effect, the $T_{cc}^+$ state matches the loosely bound
doubly charmed molecular tetraquark explanation very well. Since the
probabilities ratio for the $S-$wave $D^0D^{*+}$ and $D^+D^{*0}$
components is $2.8:1$, the isospin breaking effect does play an
important role in generating this doubly charmed molecular
tetraquark. Simultaneously, there exists a doubly charmed resonance
$T_{cc}^{\prime+}$ between the $D^0D^{*+}$ and $D^+D^{*0}$ mass
thresholds. These two doubly charmed states are the mixture of the
$D^0D^{*+}$ and $D^+D^{*0}$ components after considering the isospin
breaking effect, which satisfies
\begin{eqnarray}\label{eq11}
\left(\begin{array}{c}|T_{cc}^+\rangle\\
|T_{cc}^{\prime+}\rangle\end{array}\right) &=&
\left(\begin{array}{cc}\text{cos}\theta    &\text{sin}\theta\\
-\text{sin}\theta
&\text{cos}\theta\end{array}\right)\left(\begin{array}{c}|D^0D^{*+}\rangle\\
|D^+D^{*0}\rangle\end{array}\right).
\end{eqnarray}
By using the probability ratio for the $D^0D^{*+}$ and $D^+D^{*0}$
components in the $T_{cc}^+$ state, the mixing angle is
$\theta=\pm30.8^{\circ}$. In this scenario, the $S-$wave $D^+D^{*0}$
channel is the dominant channel for the doubly charmed
$T_{cc}^{\prime+}$ resonant tetraquark.

With the same model, the
authors of Ref. \cite{Li:2012ss} performed a coupled-channel analysis on the
$D^{(*)}D^{(*)}$ system without considering the isospin breaking
effects. They found that the OBE interactions from the $DD^*$ state with $I(J^P)=0(1^+)$ is stronger attractive than those in the
isovector state. When we consider the isospin breaking effects, there still exist two bound states, which corresponds to the $T_{cc}$ molecular state and $T_{cc}^{\prime}$ resonance, as shown in the Eq. (\ref{eq11}), the probabilities for the $D^0D^{*+}$ and $D^+D^{*0}$ components are not the same any more as in the without isospin breaking case.

The decay behavior for the exotic state is very helpful to
understand their inner structures. Very recently, several groups
discussed the strong and electromagnetic decay behavior for the
newly $T_{cc}^+$ state~\cite{Meng:2021jnw,Ling:2021bir}. The strong
and electromagnetic decay amplitudes for the $T_{cc}^+$ and
$T_{cc}^{\prime+}$ can be expressed as
\begin{eqnarray*}
\mathcal{A}_{T_{cc}^+\to D^0D^0\pi^+} &=&
\text{cos}\theta\mathcal{A}_{D^0D^{*+}\to D^0D^0\pi^+}
          +\text{sin}\theta\mathcal{A}_{D^+D^{*0}\to D^0D^0\pi^+},\\
\mathcal{A}_{T_{cc}^+\to D^+D^0\pi^0} &=&
\text{cos}\theta\mathcal{A}_{D^0D^{*+}\to D^+D^0\pi^0}
          +\text{sin}\theta\mathcal{A}_{D^+D^{*0}\to D^+D^0\pi^0},\\
\mathcal{A}_{T_{cc}^{\prime+}\to D^0D^0\pi^+} &=&
-\text{sin}\theta\mathcal{A}_{D^0D^{*+}\to D^0D^0\pi^+}
          +\text{cos}\theta\mathcal{A}_{D^+D^{*0}\to D^0D^0\pi^+},\\
\mathcal{A}_{T_{cc}^{\prime+}\to D^+D^0\pi^0} &=&
-\text{sin}\theta\mathcal{A}_{D^0D^{*+}\to D^+D^0\pi^0}
          +\text{cos}\theta\mathcal{A}_{D^+D^{*0}\to D^+D^0\pi^0},\\
\mathcal{A}_{T_{cc}^+\to D^0D^+\gamma} &=&
\text{cos}\theta\mathcal{A}_{D^0D^{*+}\to D^0D^+\gamma}
          +\text{sin}\theta\mathcal{A}_{D^+D^{*0}\to D^0D^+\gamma},\\
\mathcal{A}_{T_{cc}^{\prime+}\to D^0D^+\gamma} &=&
-\text{sin}\theta\mathcal{A}_{D^0D^{*+}\to D^0D^+\gamma}
          +\text{cos}\theta\mathcal{A}_{D^+D^{*0}\to D^0D^+\gamma}.
\end{eqnarray*}
If we only consider the contributions from the tree diagram as shown
in Ref. \cite{Meng:2021jnw,Li:2021zbw}, we can obtain $\mathcal{R}_1
= {\Gamma[T_{cc}^+\to D^0D^0\pi^+]}/{\Gamma[T_{cc}^{\prime+}\to
D^0D^0\pi^+]} = {\text{cos}^2\theta}:{\text{sin}^2\theta}=2.80$,
which explains why the significance for the $T_{cc}^{\prime+}$ state
is a little small compared to the $T_{cc}^{+}$. After neglecting the
very small contribution from the $D^{*+}\to D^+\gamma$ process,
$\mathcal{R}_2 = {\Gamma[T_{cc}^+\to
D^0D^+\gamma]}:{\Gamma[T_{cc}^{\prime+}\to D^0D^+\gamma]}
={\text{sin}^2\theta}:{\text{cos}^2\theta}= 0.35$. For the
$D^+D^0\pi^0$ final states, the decay width ratio between these two
doubly charmed tetraquark is a little complicated due to
undetermined partial decay width $\Gamma[D^{*0}\to D^0\pi^0]$.
Assuming that $\Gamma[D^{*0}\to D^0\pi^0]$ is of the same order or
even larger than $\Gamma[D^{*0}\to D^0\pi^0]$, $\mathcal{R}_3 =
{\Gamma[T_{cc}^+\to D^+D^0\pi^0]}/{\Gamma[T_{cc}^{\prime+}\to
D^+D^0\pi^0]} $ is less than 0.7. Therefore, the $D^0D^+\gamma$ and
$D^+D^0\pi^0$ decay mode shall be the prime channels to search for
the $T_{cc}^{\prime+}$ state.

{\it Summary---}At the European Physical Society conference on high
energy physics 2021, the LHCb Collaboration reported the observation
of the doubly charmed tetraquark $T_{cc}^+$ in the $D^0D^0\pi^+$
mass spectrum. Stimulated by its very near threshold property, we
perform an isospin breaking effect analysis on the
$D^0D^{*+}/D^+D^{*0}$ interactions. We adopt the OBE model and
consider the $S-D$ wave mixing effect. Our results indicate the
newly observed $T_{cc}^+$ is consistent with the
$D^0D^{*+}/D^+D^{*0}$ doubly charmed molecular tetraquark with
$J^P=1^+$, while the probabilities for the $S-$wave
$D^0D^{*+}$ and $D^+D^{*0}$ components are 72.51\% and 25.85\%,
respectively.

Using the same OBE effective potentials, we obtain a coupled
$D^0D^{*+}/D^+D^{*0}$ doubly charmed resonance $T_{cc}^{\prime+}$,
whose mass and decay width are $m=3876~\text{MeV}$, $\Gamma=
412~\text{keV}$, respectively. Based on the $T_{cc}^+$ and
$T_{cc}^{\prime+}$ being the mixture of the $D^0D^{*+}$ and
$D^+D^{*0}$ components, we further discuss their strong and
electromagnetic decay properties. Our quantitative analysis
indicates that it is a little difficult to identify the predicted
$T_{cc}^{\prime+}$ doubly charmed tetraquark in the $D^0D^0\pi^+$.
However, it is promising to search for the $T_{cc}^{\prime+}$ state
in the $D^0D^+\gamma$ and $D^+D^0\pi^0$ decay modes.

We strongly encourage our experimental colleague to pay more
attention to the structure between the $D^0D^{*+}$ and $D^+D^{*0}$
thresholds with more precise data. If this substructure can be
confirmed in the near future, it shall provide very strong evidence
of the existence of the doubly charmed molecules.

{\it Acknowledgments---}This work is supported by the National
Natural Science Foundation of China under Grants 11975033 and
12070131001, the China National Funds for Distinguished Young
Scientists under Grant No. 11825503, National Key Research and
Development Program of China under Contract No. 2020YFA0406400, and
the 111 Project under Grant No. B20063, the Fundamental Research
Funds for the Central Universities under Grants No.
lzujbky-2021-sp24. R. C. is supported by the National Postdoctoral
Program for Innovative Talent.


\begin{thebibliography}{99}

\bibitem{GellMann:1964nj}
M.~Gell-Mann, A Schematic Model of Baryons and Mesons,
{Phys. Lett. \textbf{8}, 214-215 (1964).}
\bibitem{Zweig:1981pd}
G.~Zweig, An SU(3) model for strong interaction symmetry and its
breaking. Version 1,
{CERN-TH-401.}
\bibitem{Zweig:1964jf}
G.~Zweig, An SU(3) model for strong interaction symmetry and its
breaking. Version 2,
{CERN-TH-412.}

\bibitem{LHCb:2021vvq}
R.~Aaij \textit{et al.} [LHCb],
Observation of an exotic narrow doubly charmed tetraquark,
[arXiv:2109.01038 [hep-ex]].

\bibitem{LHCb:2021auc}
R.~Aaij \textit{et al.} [LHCb],
Study of the doubly charmed tetraquark $T_{cc}^+$,
[arXiv:2109.01056 [hep-ex]].

\bibitem{Li:2012ss}
N.~Li, Z.~F.~Sun, X.~Liu and S.~L.~Zhu, Coupled-channel analysis of
the possible $D^{(*)}D^{(*)}, \overline{B}^{(*)}\overline{B}^{(*)}$
and $D^{(*)}\overline{B}^{(*)}$ molecular states, Phys. Rev. D
\textbf{88}, no.11, 114008 (2013) doi:10.1103/PhysRevD.88.114008
[arXiv:1211.5007 [hep-ph]].

\bibitem{Xu:2017tsr}
H.~Xu, B.~Wang, Z.~W.~Liu and X.~Liu, $D D^{*}$ potentials in chiral
perturbation theory and possible molecular states, Phys. Rev. D
\textbf{99}, no.1, 014027 (2019) doi:10.1103/PhysRevD.99.014027
[arXiv:1708.06918 [hep-ph]].

\bibitem{Li:2012cs}
N.~Li and S.~L.~Zhu, Isospin breaking, Coupled-channel effects and
Diagnosis of X(3872), Phys. Rev. D \textbf{86}, 074022 (2012)
doi:10.1103/PhysRevD.86.074022 [arXiv:1207.3954 [hep-ph]].

\bibitem{Gutsche:2016cml}
T.~Gutsche, M.~A.~Ivanov, J.~G.~Korner and V.~E.~Lyubovitskij,
Isospin-violating strong decays of scalar single-heavy tetraquarks,
Phys. Rev. D \textbf{94}, no.9, 094012 (2016)


\bibitem{Chen:2016qju}
  H.~X.~Chen, W.~Chen, X.~Liu and S.~L.~Zhu,
  The hidden-charm pentaquark and tetraquark states,
 {Phys.\ Rept.\  {\bf 639}, 1 (2016)}

\bibitem{Liu:2019zoy}
Y.~R.~Liu, H.~X.~Chen, W.~Chen, X.~Liu and S.~L.~Zhu, Pentaquark and
Tetraquark states,
{Prog. Part. Nucl. Phys. \textbf{107}, 237-320 (2019)}


\bibitem{Guo:2017jvc}
F.~K.~Guo, C.~Hanhart, U.~G.~Mei\ss{}ner, Q.~Wang, Q.~Zhao and
B.~S.~Zou, Hadronic molecules,
{Rev. Mod. Phys. \textbf{90}, no.1, 015004 (2018)}


\bibitem{Yan:1992gz}
  T.~M.~Yan, H.~Y.~Cheng, C.~Y.~Cheung, G.~L.~Lin, Y.~C.~Lin and H.~L.~Yu,
 Heavy quark symmetry and chiral dynamics,
  {Phys.\ Rev.\  D {\bf 46}, 1148 (1992)}
  {[Erratum-ibid.\  D {\bf 55}, 5851 (1997)].}

\bibitem{Wise:1992hn}
  M.~B.~Wise,
 Chiral perturbation theory for hadrons containing a heavy quark,
  {Phys.\ Rev.\  D {\bf 45}, R2188 (1992).}

\bibitem{Burdman:1992gh}
G.~Burdman and J.~F.~Donoghue, Union of chiral and heavy quark
symmetries,
{Phys. Lett. B \textbf{280}, 287-291 (1992)}

\bibitem{Casalbuoni:1996pg}
  R.~Casalbuoni, A.~Deandrea, N.~Di Bartolomeo, R.~Gatto, F.~Feruglio and G.~Nardulli,
 Phenomenology of heavy meson chiral Lagrangians,
  {Phys.\ Rept.\  {\bf 281}, 145 (1997)}

\bibitem{Falk:1992cx}
  A.~F.~Falk and M.~E.~Luke,
Strong decays of excited heavy mesons in chiral perturbation theory,
 {Phys.\ Lett.\  B {\bf 292}, 119 (1992)}

\bibitem{Ding:2008gr}
  G.~J.~Ding,
  Are $Y(4260)$ and {\rm$Z_2^{+}$(4250)} ${\rm D_1D}$ or ${\rm D_0D^{*}}$ hadronic molecules?
  {Phys.\ Rev.\ D {\bf 79}, 014001 (2009)}.

\bibitem{Machleidt:1987hj}
R.~Machleidt, K.~Holinde and C.~Elster, The Bonn Meson Exchange
Model for the Nucleon Nucleon Interaction,
{Phys. Rept. \textbf{149}, 1-89 (1987)}

\bibitem{Wang:2019nwt}
F.~L.~Wang, R.~Chen, Z.~W.~Liu and X.~Liu, Probing new types of
$P_c$ states inspired by the interaction between $S$-wave charmed
baryon and anti-charmed meson in a $\bar T$ doublet,
{Phys. Rev. C \textbf{101}, no.2, 025201 (2020)}

\bibitem{Isola:2003fh}
  C.~Isola, M.~Ladisa, G.~Nardulli and P.~Santorelli,
  Charming penguins in $B\to K^*\pi, K(\rho,\omega,\phi)$ decays,
  {Phys.\ Rev.\ D {\bf 68}, 114001 (2003)}.

\bibitem{Zyla:2020zbs}
 P.~A.~Zyla {\it et al.} [Particle Data Group],
 Review of Particle Physics,
{PTEP \textbf{2020}, no.8, 083C01 (2020)}.

\bibitem{Meng:2021jnw}
L.~Meng, G.~J.~Wang, B.~Wang and S.~L.~Zhu, Strong and
electromagnetic decays in the long-distance to identify the
structure of the $T_{cc}^+$, [arXiv:2107.14784 [hep-ph]].

\bibitem{Ling:2021bir}
X.~Z.~Ling, M.~Z.~Liu, L.~S.~Geng, E.~Wang and J.~J.~Xie, Can we
understand the decay width of the $T_{cc}^+$ state?,
[arXiv:2108.00947 [hep-ph]].

\bibitem{Li:2021zbw}
N.~Li, Z.~F.~Sun, X.~Liu and S.~L.~ZHu, Perfect $DD^*$ molecular
prediction matching the $T_{cc}$ observation at LHCb,
[arXiv:2107.13748 [hep-ph]].

\end{thebibliography}
\end{document}